\newcommand{\f}[1]{Fig.~\ref{#1}}
\newcommand{\eq}[1]{Eq.~(\ref{#1})}
\newcommand{\eqs}[2]{Eqs.~(\ref{#1}) and~(\ref{#2})}
\newcommand{\eqss}[3]{Eqs.~(\ref{#1}), (\ref{#2}) and~(\ref{#3})} \def\be{\begin{equation}}
\def\ee{\end{equation}}
\def\ba{\begin{array}}
\def\ea{\end{array}}
  \newcommand{\beq}{\begin{equation}}    
  \newcommand{\eeq}{\end{equation}}      
\def\bea{\begin{eqnarray}}
\def\eea{\end{eqnarray}}
  \newcommand{\bc}{\begin{center}}       
  \newcommand{\ec}{\end{center}}         
  \newcommand{\bitem}{\begin{itemize}}   
  \newcommand{\eitem}{\end{itemize}}     
  \newcommand{\bnum}{\begin{enumerate}}  
  \newcommand{\enum}{\end{enumerate}}    

\def\p{\protect}

\def\d{\partial}
\def\l({\left(}
\def\r){\right)}

\def\Bfj{B_{\rm fj}}
\def \bext{B_a}

\documentclass[aps,prb,twocolumn]{revtex4}
\usepackage{graphicx,color}
\usepackage{amsmath}
\usepackage{amssymb,bm}

\begin{document}
\bibliographystyle{apsrev}

\title{Size of flux jumps in superconducting films}

\author{D.~V. Shantsev$^{1,2}$, A. V. Bobyl$^{1,2}$, Y.~M.~Galperin$^{1,2}$, 
T.~H.~Johansen$^{1,3,}$\cite{0},
S.~I.~Lee$^{4,5}$}

\address{
$^1$ Department of Physics, University of Oslo, P. O. Box 1048, Blindern,
0316 Oslo, Norway\\
$^2$ A. F. Ioffe Physico-Technical Institute, Polytekhnicheskaya
26, St.Petersburg 194021, Russia\\
$^3$ Texas Center for Superconductivity and Advanced Materials,
University of Houston, Houston, TX 77204-5002, USA\\
$^4$ National Creative Research Initiative Center for Superconductivity, 
Department of Physics, Pohang University of 
Science and Technology, Pohang 790-784, Korea\\
$^5$ Quantum Material laboratory, Korea Basic Science Institute, 52
Yeoeun-Dong, Yusung-gu, Daejeon 305-333, Korea
}


\begin{abstract}
Magneto-optical imaging is used to visualize vortex avalanches in MgB$_2$ films at 4~K.
Avalanches ranging from 50 to 50000 vortices were detected.
The size distribution function has a clear peak whose position 
moves towards larger sizes as the applied field increases.
This field dependence as well as variation of flux density profile during an avalanche
are well described by a proposed model assuming a thermal origin of the avalanches.
The  model is based on the adiabatic approach and takes into account nonlocal
electrodynamics in thin superconductors.
The threshold field for thermal avalanches is predicted to be much smaller
than that for thick superconductors, in agreement with the experiment.
\end{abstract}

\maketitle \bc

\ec

\narrowtext
\section{Introduction}

Applying a magnetic field to a type-II superconductor results in
formation of a metastable critical state with a nonuniform flux
density that is sustained by vortex pinning. From an application
point of view the key quantity is the critical current
density that determines the flux density gradient. However, much
more information about the microscopic properties can be inferred
from the time evolution of the critical state subjected to an
external drive, e.g. a slowly increasing applied field. The
repulsive interaction between individual vortices driven through a
disordered pinning landscape may then result in avalanche
dynamics, as already observed and analyzed in a number of
experiments, see Ref.~\onlinecite{aval} for a review.
Some studies\cite{field,nowak,behnia,radovan,aeg,ernesto,NbH} have
reported a power-law distribution of avalanche sizes that is
usually interpreted as a signature of self-organized criticality
(SOC).
However, in many experiments the distribution displays a peak at
some preferred size,\cite{behnia,radovan,NbH,zieve} or sometimes
even two peaks.\cite{nowak,james,meso}

These observations have motivated molecular dynamics simulations
of the vortex motion, which predict a transition from a broad
distribution to a peaked one as the density of pinning sites
decreases.\cite{pla,olson97} However, neither the simulations nor
the SOC concept take into account thermal effects. It is well
known that vortex motion generates heat that makes pinning weaker
and facilitates further motion. This positive feedback is
responsible for large catastrophic vortex avalanches, or flux
jumps,\cite{MR} involving many millions of vortices, and usually
causes heating of the superconductor to the critical temperature.
To which degree also small-scale avalanches are affected by
thermal effects is still very unclear, and hence represents a
major problem in understanding experimental results.

To be able to distinguish between avalanches of thermal and
non-thermal origin one needs a clear theoretical prediction about
the size of thermal avalanches. Previous theoretical studies of
avalanche size considered slab superconductors placed in a
parallel magnetic field.\cite{swartz,wipf67}
This geometry is dramatically different from that in most
experiments where thin samples, often films, are placed in a
perpendicular field.
Therefore, the aim of this work is to\\
(i) present a model which enables calculation of the size of
flux avalanches in the perpendicular geometry, and\\
(ii) compare the predictions with experimental data obtained from
magneto-optical (MO) imaging  of flux dynamics in MgB$_2$
films.\\
We find that the avalanche size distribution, as well as the
detailed variation of flux density profiles during an avalanche
are in a good agreement with our model. It is shown that thermal
flux jumps can be very small -- down to
50 vortices -- if
the field is close to the threshold value.

\section{Flux jump field}

Consider a thin superconducting strip with edges located at $x=\pm
w$, thickness $d$, where $d \ll w$, and infinite in the $y$
direction. When the zero-field-cooled strip is placed in a
perpendicular magnetic field $B_{a}$, screening currents are
induced the $y$-direction. We shall assume the Bean critical
state, i.~e., $j=j_c$ in the flux penetrated region, $j_c$ being
the critical current density. Then the $z$-component of flux
density in the strip plane is\cite{zeld,BrIn} 
\be B(x)=B_c \ln{
x\sqrt{w^2-a^2}+w\sqrt{x^2-a^2}\over a\sqrt{w^2-x^2}}, \ \ a\le
|x|\le w \label{bofx} 
\ee 
and $B(x)=0$ in the central part of the
strip, $|x|<a$. Here 
\be a= {w / \cosh\left(\bext/B_c\right)},
\quad  B_c = \mu_0 j_c d / \pi \,. \label{aofb} 
\ee

\begin{figure}
\vbox{
\centerline{\includegraphics[width=8.5cm]{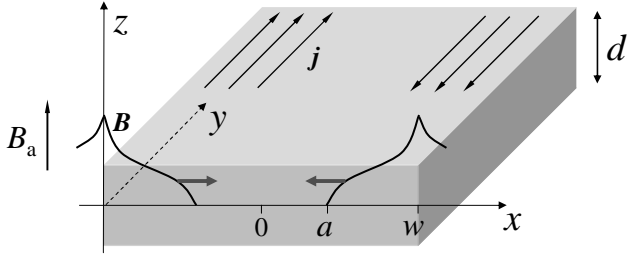}}
\vspace{0.2cm}
\caption{Superconducting strip in a perpendicular magnetic field.
\label{f_s}}}
\end{figure}

Let us assume now that a uniform fluctuation of temperature
$\delta T$ occurs. Assume also that the current-voltage curve of
the superconductor is very steep near $j_c$, so that the induced
flux motion proceeds faster than thermal diffusion. We can then 
use the adiabatic approach, i.~e., assume that the heat stays
where it is being released. The justification of this will be
discussed in detail in Sec.V. The fluctuation leads to a decrease
in the critical current density throughout the strip, by $\delta
j_{c} = |\d j_c/\d T| \delta T$. Hence, some redistribution of
flux density will occur generating the Joule heat
\be
\delta Q(x)= \int j E dt = j_c  \int\limits_a^x \delta B(x')dx', \quad x>a.\\
\label{q}
\ee
After substituting here \eq{bofx} one obtains
\begin{eqnarray}
\delta Q(x) 
= wB_c\ \left|{\partial j_c (T)\over \partial T}\right|\ \gamma(x,B_a) \ \delta T,
\label{qstrip}
\end{eqnarray}
with
\be
\gamma(x,B_a) = \int\limits_a^x{B(x')\over B_c}\ \frac{dx'}{w}+ \frac{\d a}{\d B_c}
\int\limits_a^x{\partial B(x')\over\partial a}\ \frac{dx'}{w} .
\label{fa}
\ee

Consider now the heat at the edge, 
since expectedly this is the most unstable region. 
The key quantity then is the ratio $\delta Q(w)/ \delta T$, which
should be compared with the specific heat $C$. If $C<\delta
Q/\delta T$, the generated heat cannot be absorbed, and the
temperature fluctuation will grow. Thus, one finds that the
superconductor is unstable if
\be
\beta^{-1}_{\rm} < \gamma(w,B_a) = 
\frac{B_a}{B_c}\ \tanh\frac{B_a}{B_c} -
\ln \l(\cosh \frac{B_a}{B_c}\r) 
\label{crst}
\ee
where 
\be 
 \beta^{-1}_{\rm}  \equiv \frac{\pi CT^*}{\mu_0 d w j_c^2 }, \quad 
T^* \equiv j_c \left| \frac{\d j_c}{\d T} \right|^{-1}
\label{ggl}
\ee

\begin{figure}
\centerline{\includegraphics[width=8.5cm]{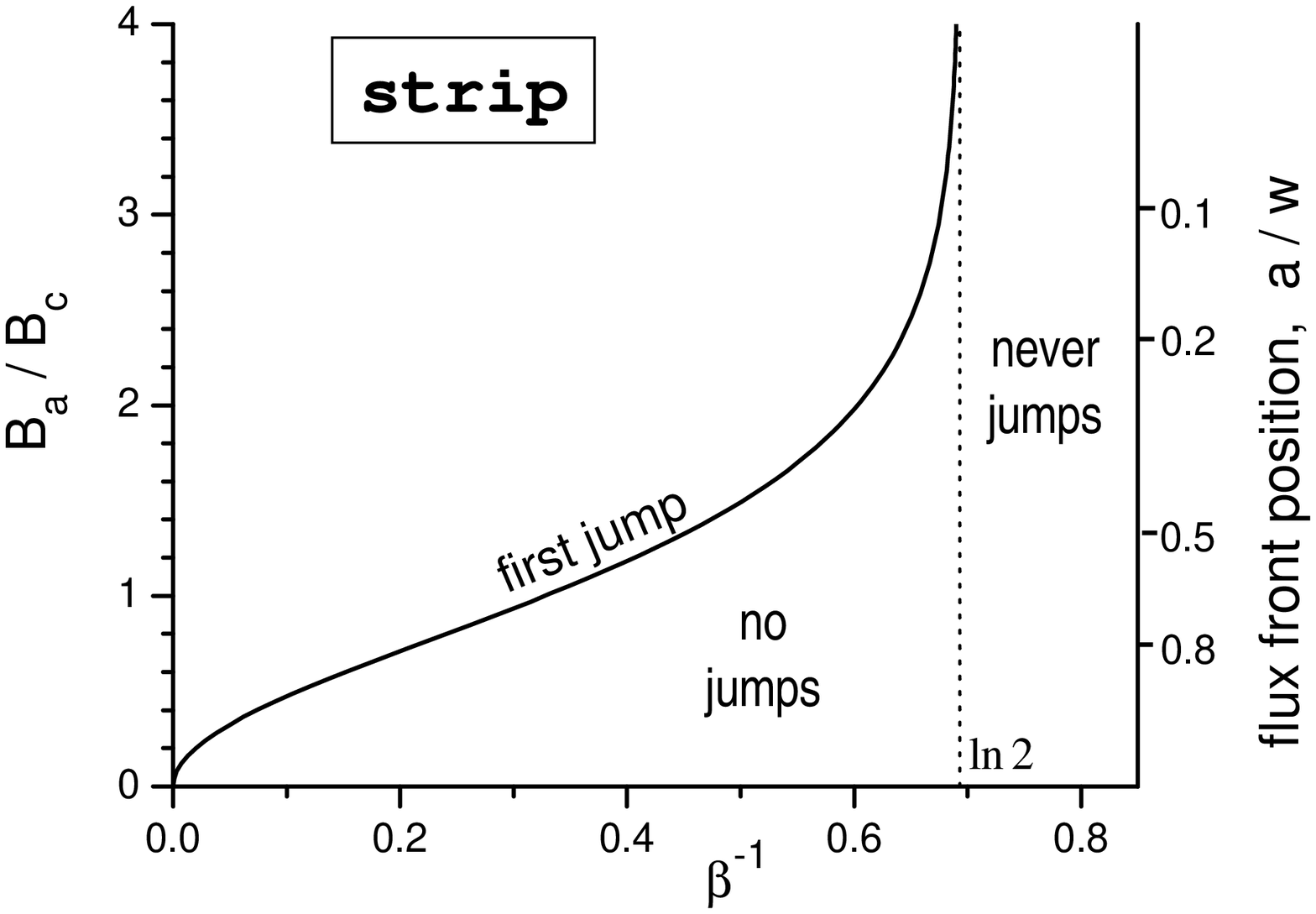}}
\centerline{\includegraphics[width=8.5cm]{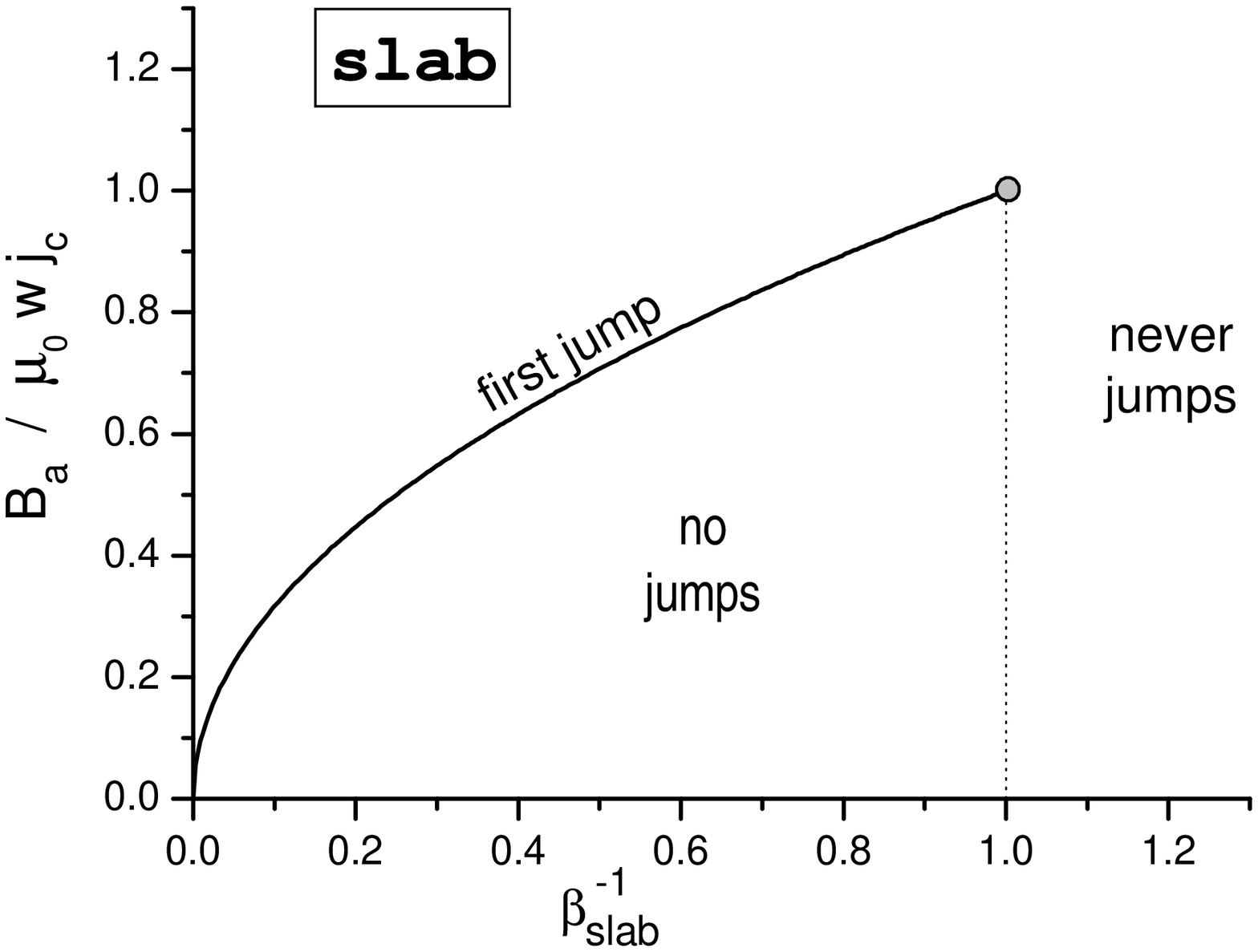}}
\vspace{0.2cm}
\caption{Stability diagrams for a strip, \p\eq{crst},
and for a slab, \eq{bfjslab}.
\label{f_str-gl}}
\end{figure}
The solution of \eq{crst} is shown graphically in \f{f_str-gl}(top). 
The superconductor is stable if the applied field is below
the so-called flux jump field $\Bfj$ represented by a solid line.
For small fields, $\gamma(w,B_a)$ grows parabolically as $\gamma \approx B_a^2/2B_c^2$, 
which allows us  
to find an explicit expression for the jump field,
\be
  \Bfj = \sqrt{ 2 \mu_0 CT^*}\ \   \sqrt{ \frac{d}{w\pi}}.
\label{bfjst}
\ee 
One can estimate from \f{f_str-gl}(top) that this expression can be used
for $\Bfj<B_c$, or, as follows from \eq{aofb}, when the flux front 
penetrates less than 40\% of the 
distance to the strip center. 
At high fields $\gamma(w,B_a)$  increases monotonously towards
the asymptotic value of $\ln 2$. 
Therefore, in a strip flux jumps never happen if 
\be
\beta^{-1}_{\rm} > \ln 2 . \label{betast} 
\ee 
It follows from \eq{ggl} that $\beta^{-1}$ is
usually a monotonously increasing function of temperature 
since $C$ grows with $T$, while $j_c$ goes down.
The temperature dependence of $T^*$ gives only a minor contribution
possibly except for $T$ very close to $T_c$.
Fig.~\ref{f_str-gl}(top) can thus be considered as an approximation for 
the stability diagram in the $B$--$T$ plane. 
It means that the condition (\ref{betast}) defines a
threshold temperature above which flux jumps are not observed no
matter how large field is applied.

For comparison, let us recall the flux jump criterion
for a slab in a parallel field.\cite{swartz,wipf67} This
problem is much simpler since the field profiles are linear, and
the result is that
flux jumps in a slab of width $2w$ can happen only when 
\be
\frac{2CT^*}{\mu_0 w^2 j_c^2 } \equiv \beta_{\rm slab}^{-1} \le 1
\label{beta} \ee and the flux jump field is \be
  B_{\rm fj}^{\rm slab} = \mu_0 w j_c \sqrt{\beta_{\rm slab}^{-1}}
  = \sqrt{2 \mu_0 C T^*} ,
\label{bfjslab}
\ee
The corresponding stability diagram is shown in \f{f_str-gl}(bottom).
It looks similar to that for a strip 
except for the existence of a well-defined point where the first-jump line ends. 
This difference stems from the fact that the applied field when the flux front 
reaches the middle of the sample, the full  penetration field, is finite for a slab, but
diverges for a thin strip.\cite{zeld,BrIn}  

Comparing \eq{bfjslab} with the corresponding \eq{bfjst} for a strip
one sees that the flux jump field for a strip is smaller than that
for a slab by a factor $\sim\sqrt{d/w}$. It can be thought of as
a ``demagnetization factor'' that characterizes the difference
between the applied field $B_a$, and the actual field at the strip
edge. This factor is essentially important
since the aspect ratio of the strip
$d/w$ is of the order of $10^{-4}$ for most
thin-film structures. Hence, thin films should
be much more unstable with respect to flux jumps than bulk
superconductors.
The threshold temperature above which flux jumps are not observed should
also be dramatically different for films and for bulk samples.
This is seen by comparing \eqs{bfjst}{beta} and noting that $\beta$
is smaller than $\beta_{\rm slab}$ by a  factor $\sim d/w$.

\section{Jump size}

When the instability condition is met, any temperature fluctuation
will trigger the development of a flux jump. In this section we
will focus on the final state that the superconducting strip
reaches after the jump.
Using again the adiabatic approach,
we find that the final temperature distribution $T(x)$
satisfies the following equation, which is obtained by
integration of \eq{q},
\be
\int_{T_0}^{T(x)} \frac{C(T)\
dT}{j_c(T)}  = \Phi(x) . \label{act}
\ee
Here $T_0$ is the initial
temperature before the jump, and 
\be
\Phi(x) = \int_0^x
\left[B(x')-B_0(x')\right] dx', \label{acf}
\ee 
 is the flux per unit
length that has passed through the point $x$ during the whole
course of the jump.
 $B_0(x)$ is
the initial flux distribution. Again we employ the Bean critical
state model,
so that $B_0(x)$ is given by \eq{bofx} with $j_c=j_c(T_0)$.
Similarly, in the final state one has $j(x)=j_c[T(x)]$ for all $x$
where $B(x) \neq 0$. To complete the set of equations, the flux
and current density distributions are connected via the
Biot-Savart law,
\be
B(x) = B_a - \frac{\mu_0 d}{\pi} \int_0^w
\frac{j(x') }{x^2-x'^2}\ x'\ dx', \label{acb}
\ee
(where the
symmetry $B(x)=B(-x)$ was taken into account).

In their solution of the analogous case of a slab in a parallel
field Swartz and Bean\cite{swartz} called the final state after a
flux jump
{\em the adiabatic critical state}. The adiabatic critical state
for a thin strip is therefore defined by \eqss{act}{acf}{acb}. We
shall solve these equations numerically for the most
commonly used $T$ dependence of $j_c$ and $C$, namely the simple
linear and cubic forms,
\be
  j_c(T) = j_{c0} (1- T/T_c), \quad  C(T) = C(T_c) (T/T_c)^3.
\ee 
Note that as the temperature increases during a jump, the heat
capacity
also grows, resulting in a stabilization of the flux jumps thus
limiting their size.

The solution of \eqss{act}{acf}{acb} is not necessarily unique.
To ensure that we find the proper one, we build the solution
incrementally simulating the evolution of the system during a flux jump.
We start from a very small uniform temperature fluctuation $\delta T$ and calculate
the electric field $E\propto j(x)-j_c[T(x)]$ everywhere where $j>j_c$.
The field and temperature variations are found as $\delta B = \d E/\d x$ and
$\delta T = jE/C(T)$. Then we recalculate $E$, and continue until
$j\le j_c$ everywhere in the strip. The $B$ and $T$ distributions obtained
by this procedure will satisfy  \eqss{act}{acf}{acb} .

Shown in \f{f_ht} are flux density and temperature profiles in the
adiabatic critical state found by solving \eqss{act}{acf}{acb}.
The flux profiles extend deeper into the strip than the initial
critical-state profiles, which are shown by the dashed
curves. The flux density $B(x)$ inside the superconductor has
mainly increased, indicating penetration of additional flux into
the strip during the jump.
In a small region near the edge, and also on the outside, $B(x)$
has decreased in order to conserve the total amount of flux. The
temperature has increased throughout the region where the flux
motion took place. The single point where $B(x)$ did not change is
obviously the point through which the maximum amount of flux
$\Phi(x)$ has passed. Therefore it also shows the
maximal temperature increase.

The jump size can be characterized by the total amount of flux
arriving into the strip during the jump, $\Phi(w)$. Shown in
\f{f_f} is $\Phi(w)$ plotted as a function of the applied field.
For small $B_a$ there are no jumps, in full agreement with the
instability criterion \eq{crst}. As $B_a$ increases above the
threshold, the jump size grows too, eventually reaching a
saturation value at high fields.

\begin{figure}
\vbox{
\centerline{\includegraphics[width=8.5cm]{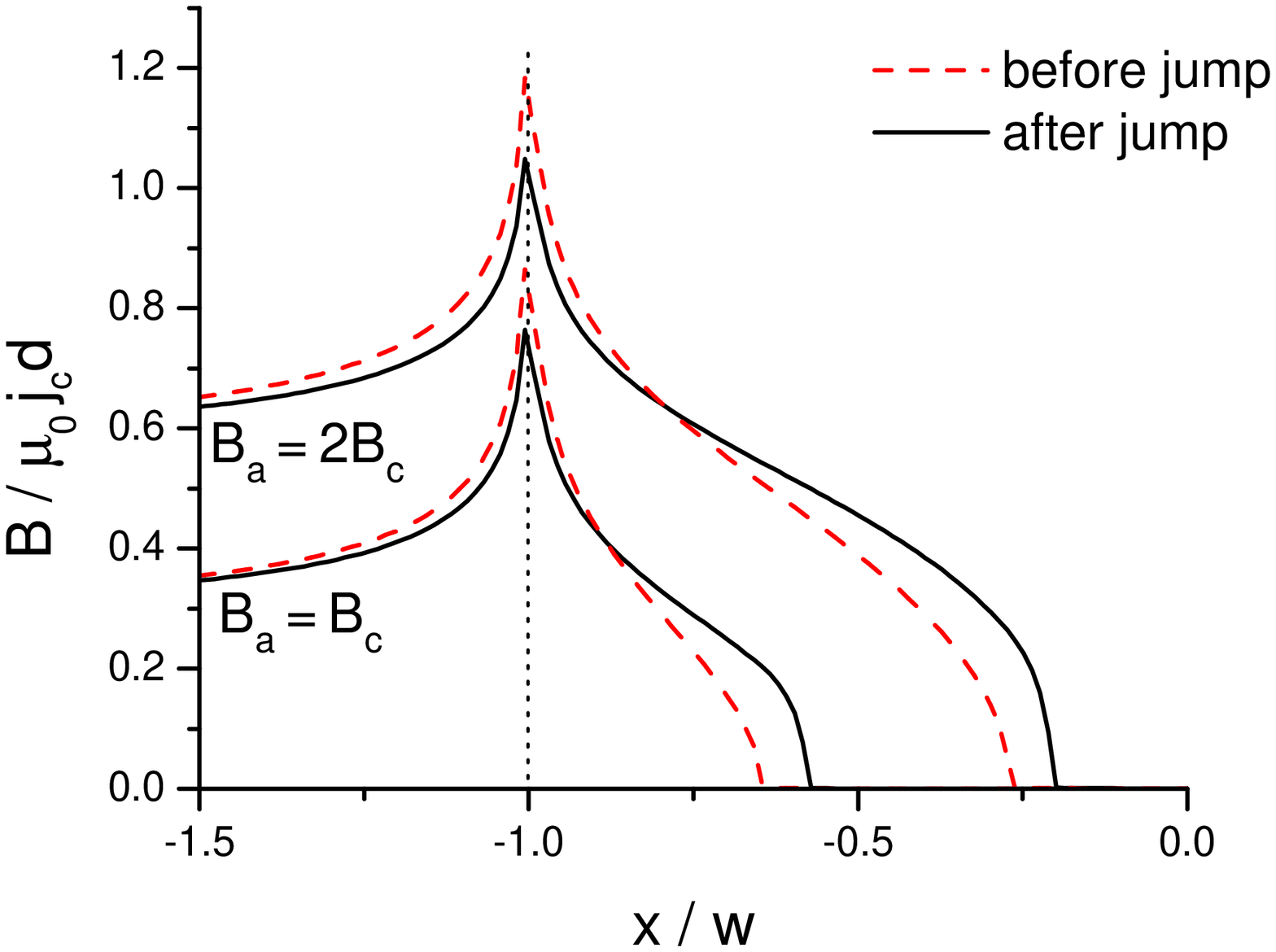}}
\centerline{\includegraphics[width=8.5cm]{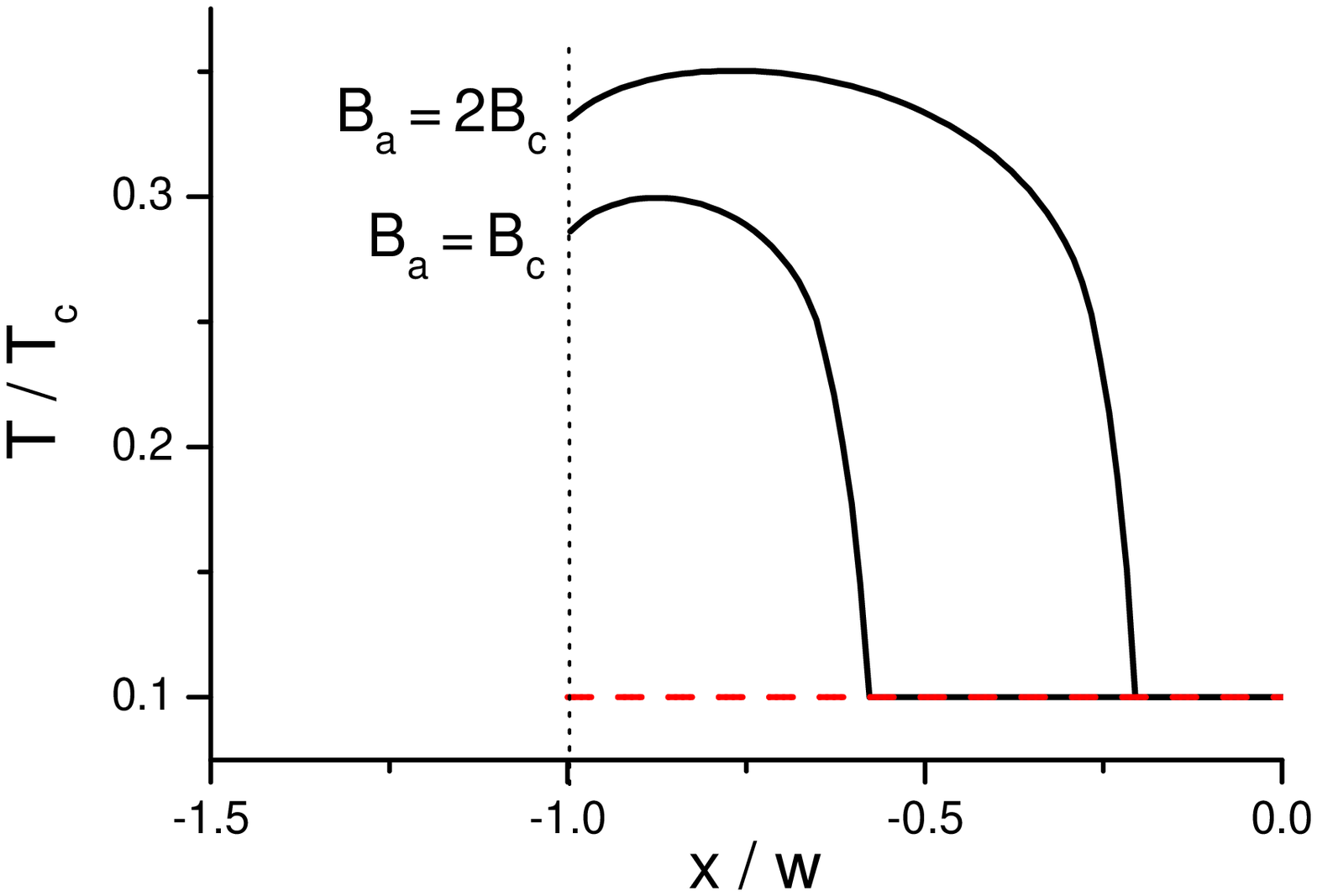}}
\caption{Flux density and temperature distributions before the jump (critical state)
and after the jump (adiabatic critical state) for two applied magnetic fields,
$\beta^{-1}(T_c/4)=\ln 2$, and $T=0.1T_c$.
\label{f_ht}}}
\end{figure}

\begin{figure}
\centerline{\includegraphics[width=8.5cm]{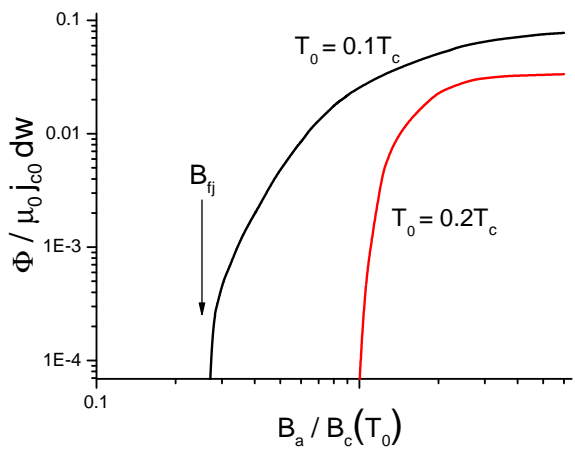}}
\caption{The flux jump size as a function of applied magnetic field
for different $T_0$ and $\beta^{-1}(T_c/4)=\ln 2$. 
\label{f_f}}
\end{figure}


\section{Experiment}

Flux jumps were studied in
films of MgB$_2$ fabricated on Al$_2$O$_3$ substrates using
pulsed laser deposition.\cite{sample}
The samples had thickness of 400~nm and lateral dimensions
4.8$\times$4.8~mm$^2$. The films had a high degree of $c$-axis alignment
perpendicular to the plane, and showed a sharp superconducting
transition at $T_c=39$~K.

The flux density distribution
was visualized using MO imaging based on the Faraday effect
in ferrite garnet indicator films. For a recent review of the
method, see Ref.~\onlinecite{MO}, and a description of our setup is found
elsewhere.\cite{joh96} The sample was glued with GE varnish to the
cold finger of the optical cryostat, and a piece of MO indicator
covering the sample area was placed loosely on  top of the
MgB$_2$ film. The grey levels in the MO images were converted
to magnetic field values using a position-dependent calibration matrix.

Flux penetration was studied in zero-field-cooled films subjected to
a slowly increasing perpendicular magnetic field $B_a$.
The field ramp rate was chosen sufficiently slow, typically 0.01~mT/s, to give
results that were rate independent.
Shown in \f{f_mo}(a) is a MO image of a region near the film edge
at $B_a=7.15$~mT.
The flux density is maximal at the film edge 
which is located near the bottom of the image.
Most part of the film remains in the Meissner state which appears black
on the image.  The flux has here penetrated only a distance of $\sim 100$~$\mu$m into film,
and the flux front is seen to be strongly nonuniform.
Such a nonuniformity is rather typical for superconducting films.
It reflects the presence of defects at the edge that facilitate vortex entry.

To detect and quantify flux jumps
we shall analyze difference images of the type shown in \f{f_mo}(b-f).
They are obtained by subtraction
of subsequent MO images taken with 2~s  intervals which implies $\Delta B_a=0.02$~mT.
The medium grey color here corresponds to unchanged flux density, while brighter areas
show where  $B$ has increased. In all these images
one can clearly see bright spots on the grey background, and notice that
their position always changes. 
These spots indicate flux jumps -- an abrupt increase
of flux density within a localized area.
One can integrate the increase of flux density over the jump area
to find the total amount of flux arrived there.
E.~g., for the jump shown in (g) it equals 900 flux quanta, while
the average increase of $B$ within the encircled region is 2.2~mT.
The duration of one jump is shorter than 0.1~s since the human eye could not follow its evolution.
We emphasize  that shown in \f{f_mo}(b-f) is not the full series of difference images.
Approximately 50\% of images were omitted because they do not display any jumps.

\begin{figure}
\vbox{
\centerline{\includegraphics[width=7.5cm]{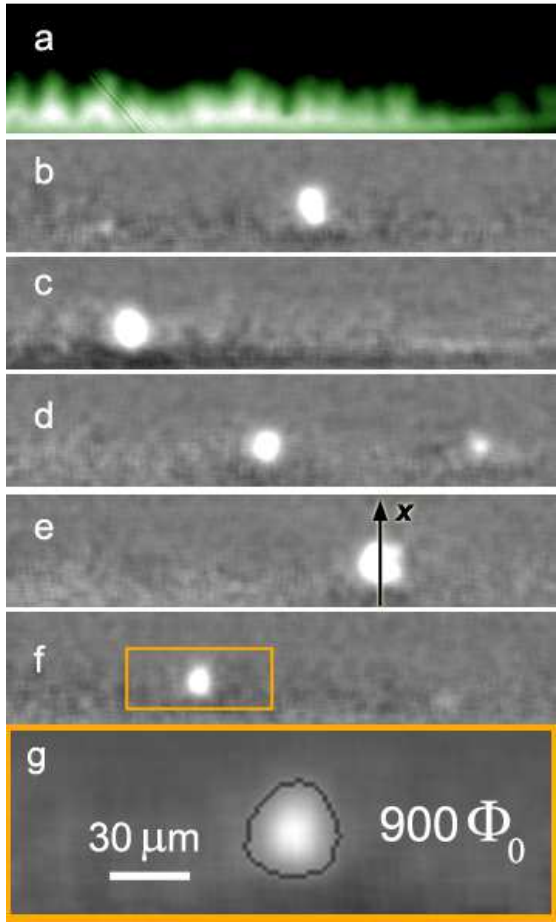}}
\caption{(a) MO image showing flux distribution near the edge of MgB$_2$ film
for $B_a=7.15$~mT at 4~K. (b-f)
Difference images obtained by subtracting two subsequent MO images taken
with field interval $\Delta B_a=0.02$~mT display flux jumps (bright spots).
(g) A blow-up of the marked rectangular area on (f);
the flux arrived within the outlined jump area is 900$\Phi_0$, 
$\Phi_0=h/2e$ is the flux quantum.
\label{f_mo}}}
\end{figure}

Shown in \f{f_prof} are profiles of flux density 
before and after flux jumps. They are recorded along the $x$ axis 
perpendicular to the film edge, as shown in \f{f_mo}(e).
The curves are very similar to the theoretical ones presented in \f{f_ht}(top).
Again we see an increase of $B$ inside the film that is especially large close
to the flux front, and a slight decrease outside.

\begin{figure}
\vbox{
\centerline{\includegraphics[width=8cm]{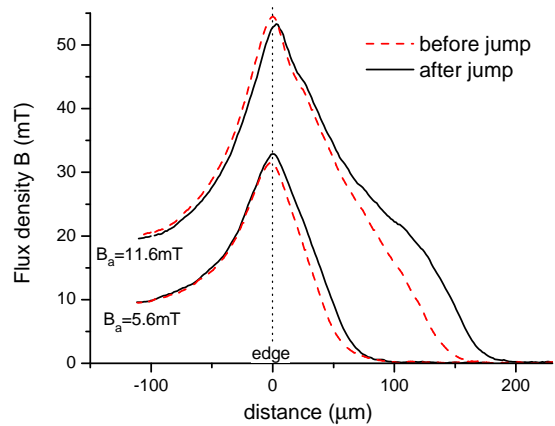}}
\caption{Profiles of flux density near the edge of MgB$_2$ film
before and after flux jump; the jump sizes are 7500$\Phi_0$ for 5.6~mT and 31000$\Phi_0$ for 11.6~mT.
\label{f_prof}}}
\end{figure}

Let us now analyze quantitatively the size distribution of flux jumps.
We ramped the applied field from zero to 23~mT at 3.6~K and
kept record of all flux jumps that took place within
our observation window of 0.7~mm length along the film edge.
The results are shown in \f{f_many} where every jump is represented by one symbol.
The first jumps are detected around $B_a=3$~mT and are small, i.e. containing 100 or less vortices.
As $B_a$ increases, the average jump size steadily grows although
with a large scatter of data.
At $B_a=14.5$~mT we observed a jump of clearly different type.
It is much larger than all the preceding
jumps, extends on a distance of $\sim \! 0.5$~mm, and
has a distinct finger-like pattern. Another jump of the same type
was observed at $B_a=17.5$~mT. Similar jumps, often called dendritic, have been reported earlier
in films of MgB$_2$\cite{epl,prb,phc} and several other superconductors,
in particular, YBa$_2$Cu$_3$O$_{7-x}$,\cite{leiderer,bolz03} Nb,\cite{duran}
 Nb$_3$Sn,\cite{rudnev} YNi$_2$B$_2$C,\cite{jooss} Pb,\cite{menghini} and NbN.\cite{nbn}
One sees from \f{f_many} that each dendritic jump is followed
by an ``empty'' interval of $B_a$ free of the small jumps.
Obviously, such a massive flux intrusion significantly reduces the ``magnetic pressure''
on distances of the order of dendrite length.
A considerable increase of applied field is then needed to build up this pressure again
in order to trigger new jumps.

Apart from the region $B_a>14$~mT affected by dendritic jumps, the observed
$\Phi(B_a)$ dependence is in a good agreement with the curve
predicted for $T=0.1T_c$, see \f{f_f}.
Both the experimental and theoretical $\Phi(B_a)$ curves have a 
steep initial slope and tend to saturation 
at $B_a\sim 5 B_{fj}$. Note that this qualitative agreement is achieved without any
fitting parameters. The only parameter in the model, $\beta$,  
is fixed by
the experimental fact that the jumps do not occur above 10~K$\approx T_c/4$.
Hence, according to \eq{betast}, we have chosen $\beta^{-1}(T_c/4)=\ln 2$.

The large scatter of the measured jump sizes is probably due to sample
inhomogeneities that result in a very nonuniform penetration depth along the flux front that
is seen in \f{f_mo}(a).
Another factor increasing the scatter is the influence of preceding avalanches that
can affect the initial state for the following flux jumps.
To quantify the scatter of the jump sizes we collect statistics from the
whole sample and plot the corresponding
distribution functions in \f{f_disex}.
The distribution strongly depends on the applied field $B_a$, as expected from \f{f_many}.
Despite the distributions are rather broad, the increase of the average jump size as $B_a$ changes from 4 
to~10~mT is quite clear.
Strictly speaking, the distribution function for $B_a=4$~mT might not be very accurate
since we could have missed some jumps smaller than 50$\Phi_0$ which is our resolution limit. However,
for $B_a=10$~mT there is a definite peak in the range 1000-4000$\Phi_0$.
This rules out any possibility for SOC behavior characterized by a power-law size distribution.
For larger $B_a$ a distinct second peak appears in the size distribution due to dendritic avalanches.

\begin{figure}
\vbox{
\centerline{\includegraphics[width=8.5cm]{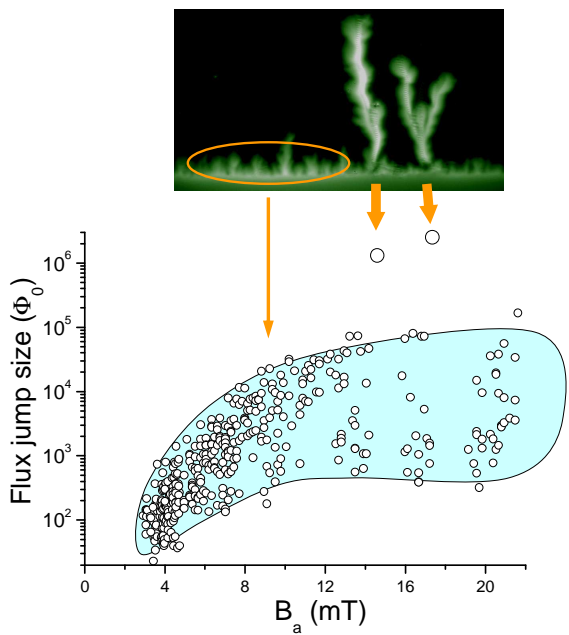}}
\caption{Statistics of flux jump sizes in MgB$_2$ film: every symbol corresponds to one jump.
\label{f_many}}}
\end{figure}

\begin{figure}
\vbox{
\centerline{\includegraphics[width=8cm]{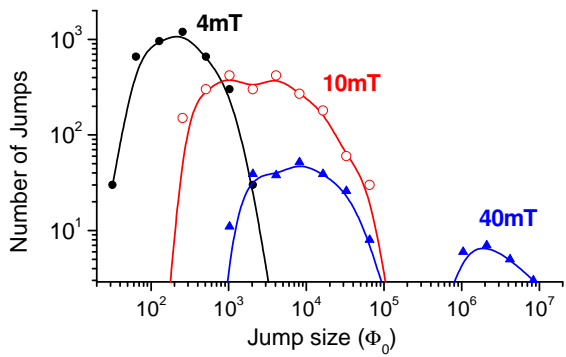}}
\caption{Size distribution flux jumps.
The statistics is collected over all jumps that occurred in MgB$_2$ film
within 3~mT intervals of applied field centered around $B_a$=4, 10 and 40~mT.  
\label{f_disex}}}
\end{figure}

To study the reproducibility of the observed jumps we
carry out several identical experiments under exactly the same conditions.
In order to visualize the difference between the flux distributions
measured in three experiments, the three MO images were coded as red, blue and green,
and then merged together, see \f{f_irrep}(top). The image seems almost
perfectly black-and-white demonstrating that the flux distribution is very well
reproducible from one experimental run to another.
However, if one combines in a similar way three {\em difference} images,
the resulting image is full of colors, see \f{f_irrep}(bottom).
This means that positions and sizes of flux jumps that took place
as the applied field was increasing from 8.5 to 8.7~mT are not reproducible.
For example, the jump (a) occurred only in one experiment (green).
The jump (b) occurred in all three experiments, however its
position and strength were not the same. It was the strongest
in the ``blue'' experiment, but somehow weaker and shifted to the left in the ``red''
experiment, and to the right in the ``green'' one.
Only the jump (c) was fairly reproducible as seen by a purely white region.
In this case the place of the jump was predefined by a defect in the film.
The presence of this defect can be immediately noticed by a reproducible peculiar
feature in the flux distribution.
We conclude therefore that in the absence of pronounced defects
a unique and irreproducible sequence of flux jumps in every experiment
produces a well reproducible final flux distribution.

\begin{figure}
\vbox{
\centerline{\includegraphics[width=8cm]{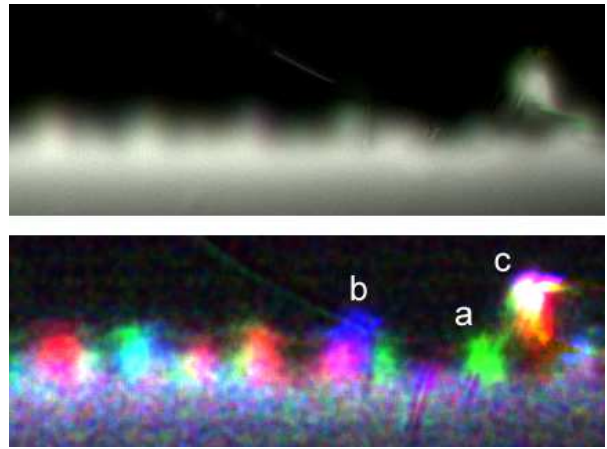}}
\caption{Irreproducibility of flux jumps.
Top: Three MO images taken at $B_a=8.7$~mT in
identical experiments are colored as red, blue and green and merged together.
Bottom: Similar color merging for difference images that show flux jumps
that took place between $B_a=8.5$ and 8.7~mT.
Positions and sizes of flux jumps are unique for every experiment, though
the final flux distribution is well reproducible.
\label{f_irrep}}}
\end{figure}

\section{Discussion}

Most predictions of the proposed adiabatic model are in a good agreement with
the experimental results, namely, (i) the field dependence of the avalanche size,
$\Phi(B_a)$ shown 
in Figs.~\ref{f_f} and~\ref{f_many}, and (ii) the flux density profiles 
before and after a jump shown in
Figs.~\ref{f_ht} and~\ref{f_prof}.
It suggests that the small vortex avalanches we observe in the MgB$_2$ film
are of thermal origin. 
Another argument in favor of the thermal origin is the presence of 
huge dendritic jumps in the same film and in the same temperature range.
The thermal origin of the dendritic jumps is now well established 
both experimentally (they are suppressed by a contact with metal\cite{phc,gold})
and theoretically.
\cite{epl,rakh04,aranson04,biehler,dima}

Coexistence of rather round and elongated dendritic jumps, both of thermal
origin, can be understood within the recent theoretical models.\cite{rakh04,aranson04,dima} 
They predict three possible situations depending on the parameters 
of superconductor and the background electric field $E$: 
(i) stability, (ii) instability leading to
uniform jumps, and (iii) instability leading to dendritic jumps.
A transition from uniform to dendritic jumps should occur
when $E$ exceeds a threshold value. 
Experimentally, $E$ is created by ramping magnetic field, and grows with $B_a$. Hence,
the appearance of dendritic jumps at larger $B_a$ seen in \f{f_many} perfectly agrees with
the theories.\cite{crit-scale}

An important question now is 
whether the proposed adiabatic model predicting a jump size can be applicable to 
jumps of any shape.
The model considered an ideal uniform film where a jump
occurs uniformly along the flux front.
The real sample is nonuniform, and the instability condition cannot be
simultaneously met along the whole length of the front. 
Hence, every jump
remains localized in the $y$ direction and eventually
acquires an approximately round shape, or grows into a long dendrite.
We made a rough estimate of the flux density increase due to 
current redistribution corresponding to a uniform and to a round jump  
schematically shown in \f{f_round}. In the jump center the 
results are different only by a numerical factor of $\pi/2$.
The resulting temperature rise is, in the adiabatic approach, determined only by the local
flux motion and, hence, independent of the jump shape. 
This justifies the use of the proposed model to predict the size of round jumps.
However, if a jump acquires a dendritic shape, 
the current distribution changes so significantly that  
the ``uniform'' model fails. Strong bending of the current  
flow around the jump area creates additional acceleration of  
the jump development. As a result, the size of dendritic jumps 
is usually limited only by the sample dimensions.


\begin{figure}
\vbox{
\centerline{\includegraphics[width=7.5cm]{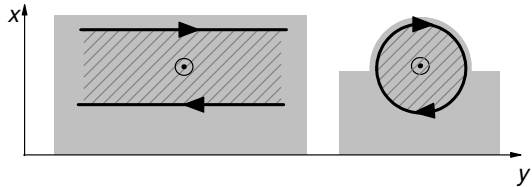}}
\caption{Schematics of current redistribution during a uniform (left)
and a round (right) flux jump. The flux penetrated area is grey,
dashed is the region where it has increased. Magnetic fields generated
by two infinite wires (a) and by a current loop (b) differ only by a factor of $\pi/2$.
\label{f_round}}}
\end{figure}

Our data show that the small jumps emerge as precursors of dendritic jumps
when the applied magnetic field is increasing from zero.
We believe that this is a general rule and that small jumps occur
everywhere where dendritic jumps do.
This explains the presence of two characteristic jump sizes reported
in a number of experiments.\cite{nowak,james,meso}
On the other hand,
one can imagine a situation when small flux jumps do occur, while the dendritic jumps do not.
Then, even at very large applied fields there will be a special type of critical state which
is formed by repeated local flux jumps. This was possibly the case
in experiments on vortex avalanches where a peaked size distribution was
measured.\cite{behnia,radovan,zieve,nowak,james}

The adiabatic approach used in our model implies that magnetic flux is moving faster than heat.
This is more than likely for the case of dendritic jumps where the flux
is moving at velocities as high as 10-100~km/s.\cite{leiderer,bolz03}
Moreover, the very fact that a jump acquires a dendritic shape
implies, according to recent models,\cite{rakh04,aranson04,dima} that the adiabatic approach holds.
At the nucleation stage of flux jump the flux motion must be relatively slower, but even here
the adiabatic approach seems to work well. Indeed,
it predicts the flux jump field, \eq{bfjst}, $B_{fj}=1.3$~mT
in a reasonable agreement with experiment
(for $C(4$K$)=0.3$~kJ/Km$^3$,
Ref.~\onlinecite{walti}).
Had we been far from the adiabatic limit, the
adiabatic model would have underestimated the flux jump field
by a large ratio of thermal and magnetic diffusivities.\cite{MR,aboutMB}

The total amount of flux that entered the strip during
field increase from 10 to 11~mT can be
found by subtracting two corresponding MO images.
This amount turns out to be 2 times larger than the sum over
all flux jumps detected in the specified range of $B_a$.
It means that only one half of flux arrived into the superconductor
via flux jumps. The other half arrived via gradual penetration
(or very small jumps that are below our resolution).
This is not surprising. The adiabatic critical state established after
every jump is characterized by less steep flux gradients than
the original critical state. Therefore every jump should be followed
by a relatively quiet period when increasing $B_a$ results in a gradual
penetration. The flux gradients build up during this period
bringing the system to the instability threshold again.
We found that flux jumps in the same area occur
with an interval of  $\Delta B_a\approx $0.5~mT.\cite{meso}

Our analysis suggests that thermal flux jumps in superconducting films
can be virtually microscopic, down to at least 50 flux quanta.
The thermal effects should therefore be carefully considered when analyzing
statistics of vortex avalanches even for small avalanche sizes.
This is especially important for Hall probe measurements that cannot access the
spatial pattern of avalanche and always underestimate its actual size.

In conclusion, we propose an adiabatic model for flux jumps in thin
superconductors. We find the flux and temperature distributions in the
adiabatic critical state after a jump, the flux jump field, and the
threshold temperature above which jumps disappear.
We find how jump size depends on applied field and demonstrate, also experimentally,
that thermal jumps can be virtually microscopic.
In this sense they do not destroy the critical state,
rather they present the mechanism of its formation.

\acknowledgements
The financial support  by  FUNMAT/UiO and the 
Research Council of Norway, Grant. No. 158518/431 (NANOMAT)
is gratefully acknowledged.
The authors wish to thank
E. Altshuler, R. Wijngaarden, V. Vlasko-Vlasov for fruitful discussions and O. Shalaev
for help.

\widetext

\begin{references}
\bibitem{crit-scale}
In real samples development of dendritic jumps can be hindered
by nonuniformity of the flux front. Therefore another criterion can be of practical
importance: a jump
needs to exceed the typical scale of flux front roughness to get a chance to grow into a dendrite.
This criterion is also in agreement with experiment because the jump size
grows with $B_a$ too.
\bibitem{aboutMB} The flux jump field for a thin strip found in Ref.~\onlinecite{MintsBr}
in the dynamic limit is much higher than that in the adiabatic limit found in this work.
\end{references}
\end{document}